# Implementation of CRISP Methodology for ERP Systems

S.Hanumanth Sastry
Department of CS & SE
Andhra University
Visakhapatnam – 530003, India
E-Mail: hanusistla@gmail.com

Prof.M.S.Prasada Babu
Department of CS & SE
Andhra University
Visakhapatnam – 530003, India
E-Mail: msprasadbabu@yahoo.co.in

**Abstract—** ERP systems contain huge amounts of data related to the actual execution of business processes. These systems have a particular way of recording activities which results in an unclear display of business processes in event logs. Several works have been conducted on ERP systems, most of them focusing on the development of new algorithms for the automatic discovery of business processes. We focused on addressing issues like, how can organizations with ERP systems apply process mining for analyzing their business processes in order to improve them. The data handling aspect of ERP systems contrasts with those of BPMS or workflow based systems, whose systematical storage of events facilitates the application of process mining techniques. CRISP-DM has emerged as the de facto standard for developing data mining and knowledge discovery projects. Successful data mining requires three families of analytical capabilities namely reporting, classification and forecasting. A data miner uses more than one analytical method to get the best results. The objective of this paper is to improve the usability and understandability of process mining techniques, by implementing CRISP-DM methodology for their application in ERP contexts, detailed in terms of specific implementation tools and step by step coordination. Our study confirms that data discovery from ERP system improves strategic and operational decision making.

## I.    INTRODUCTION

Data mining can be used to automatically determine significant patterns and hidden associations from large amounts of data. Data mining provides insights and correlations that had formerly gone unrecognized or been ignored because it had not been considered possible to analyze them [1]. Generally speaking, Functional requirements and Business rules cannot be ported from one ERP implementation to another because local operations have highly specific [2] and customized procedures for operating that equipment [3] [4]. ERP systems implement highly customized solutions to meet specific business requirements of enterprises. Since each enterprise has different data mining requirements, it is not possible to deliver fixed models for producing prediction results. Data mining professionals are challenged with developing different models to meet such business requirements which help in decision making.  Data extracted from ERP system can be mined to answer business questions like – differential product pricing to match customer profile, customer churn, cross selling potential of new products, cash flow analysis etc. ERP can play an essential role in Driving accurate and fast decisions (product profitability, procurement spend) with consistently defined data [3]. The benefits from ERP implementation can be measured both in qualitative & quantitative terms like- efficient business processes, enhanced customer service, reduced costs, improved productivity, accelerated transaction time, workflow management and reduction in the number of credit management errors.  [4]. In this paper we have discussed how CRSIP-Data Mining methodology can be implemented on ERP system data of a large manufacturing enterprise, where SAP is the ERP solution provider [5]. The rest of paper is organized as follows – Part II discusses Data in ERP Systems, Part III outlines CRISP-DM methodology, Part IV discusses Implementation of CRISP-DM, Part V specifies Implementation steps, Part VI discusses Results and finally Part VII provides Conclusions on paper.

## II.    DATA IN ERP SYSTEMS

ERP systems provide an increased level of integration to support core business processes and is an amalgamation of three most important components - Business Management Practices, Information Technology and Specific Business Objectives. Till recent past operational and transactional needs and not information were the focus of most ERP implementations. At the core of ERP is a well managed centralized data repository which acquires information from and supply information into the fragmented applications operating on a universal computing platform. Information in large business organizations is accumulated on various servers across many functional units and sometimes separated by geographical boundaries [4]. Such information islands can possibly service individual organizational units but fail to enhance enterprise wide performance, speed and competence. In these cases, it is sometimes necessary to gather the scattered data into a single database called a Data Warehouse (DW), before submitting it to data mining activity. The key objective of an ERP system is to integrate information and processes from all functional divisions of an organization and merge it for effortless access and structured workflow [5]. The integration is typically accomplished by constructing a single database repository that communicates with multiple software applications providing different divisions of an organization with various business statistics and information.

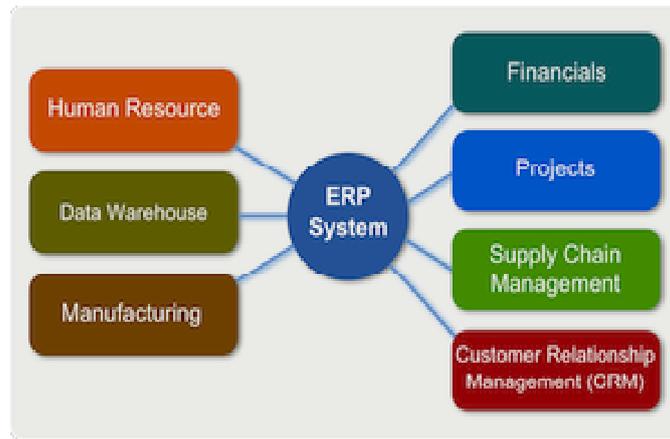

Figure 1. Integration of different modules in ERP systems

The Principal aim of Data Mining working in ERP Systems is to process data in Business Information Warehouse for automation of decision-making and forecast processing. SAP offers a complete data warehousing environment that simplifies the most challenging task in building a data warehouse – the data capture from ERP applications and building closed-loop feedback mechanisms with business critical applications [7] [8]. For our study, Cash Flow process data of Finance Module is used for implementing CRSIP methodology and extract useful information. The Cash flow statement consists of ERP Data from Operations, Investments, Loans, Payments & Receipts from branches and Special projects domains.

### III. CRISP-DM METHODOLOGY

CRISP (Cross Industry Standard Process for Data Mining), is a data mining process model that describes commonly used approaches that expert data miners use to tackle business problems [5]. It borrowed ideas from the most important pre-2000 models and is the groundwork for many later proposals. The CRISP-DM 2.0 Special Interest Group (SIG) was set up with the aim of upgrading the CRISP-DM model to a new version better suited to the changes that have taken place in the business arena since the current version was formulated [6]. The CRISP-DM methodology is described in terms of a hierarchical process model, consisting of sets of tasks described at four levels of abstraction (from general to specific): phase, generic task, specialized task, and process instance. CRISP-DM is divided into six phases to be carried out in a DM project as shown in figure 2. Implementation details in each phase are also given in table I & CRISP-DM objectives are in table II.

### IV. IMPLEMENTATION OF CRISP-DM METHODOLOGY

Creation of the model is generally not the end of the project. Even if the purpose of the model is to increase knowledge of the data, the knowledge gained will need to be organized and presented in a way that the customer can use it. The methodology we have followed for Cash Flow data mining is shown in figure 3 [5] [6] [7].

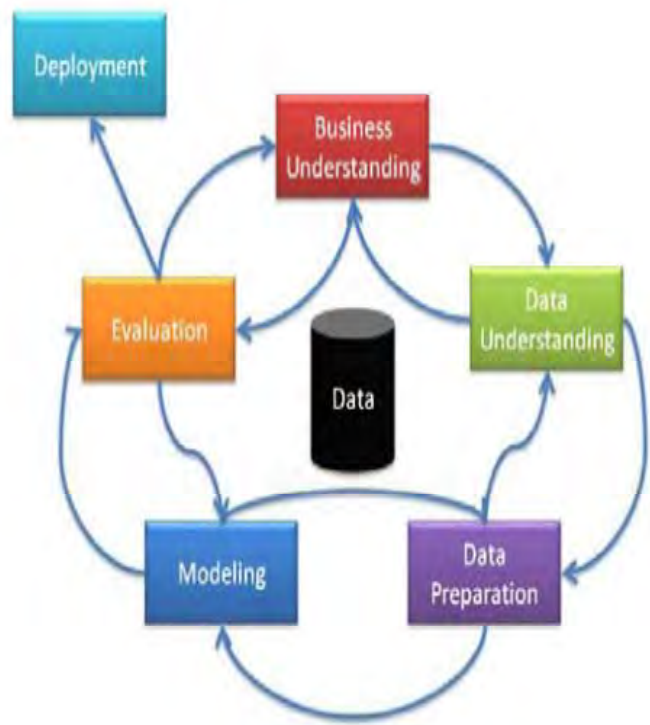

Figure 2.    Phases in CRSIP-DM Methodology

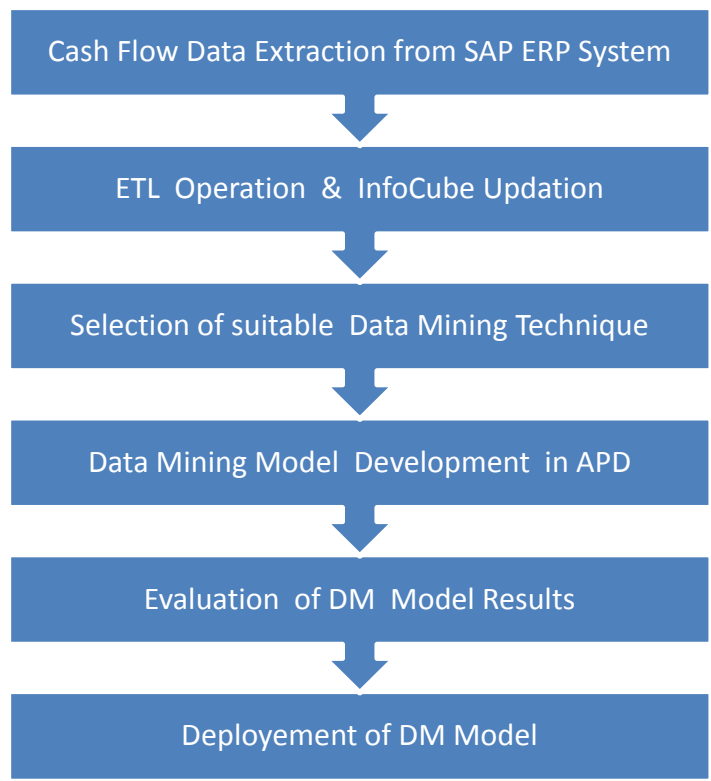

Figure 3.    CRISP-DM Methodology for Cash Flow Data Mining

TABLE I.          IMPLEMENTATION DETAILS IN EACH PHASE

| Phase in CRISP-DM | Implementation Details |
|---|---|
| Business Understanding | Understanding the project objectives and requirements from a business perspective, and then converting this knowledge into a data mining problem definition. A preliminary plan is designed to achieve the objectives. |
| Data Understanding | It starts with an initial data collection and proceeds with activities in order to get familiar with the data, to identify data quality problems, to discover first insights into the data, or to detect interesting subsets to form hypotheses for hidden information. |
| Data Preparation | It covers all activities to construct the final dataset (data that will be fed into the modeling tool(s)) from the initial raw data. Data preparation tasks are likely to be performed multiple times, and not in any prescribed order. Tasks include table, record, and attribute selection as well as transformation and cleaning of data for modeling tools. |
| Modeling | In this phase, various modeling techniques are selected and applied, and their parameters are calibrated to optimal values. Typically, there are several techniques for the same data mining problem type. Some techniques have specific requirements on the form of data. Therefore, stepping back to the data preparation phase is often needed. |
| Evaluation | At this stage in the project we would have built a model (or models) that appear to have high quality, from a data analysis perspective. Before proceeding to final deployment of the model, it is important to thoroughly evaluate the model, and review the steps executed to construct the model, to be certain it properly achieves the business objectives. A key objective is to determine if there is some important business issue that has not been sufficiently considered. At the end of this phase, a decision on the use of the data mining results should be reached. |
| Deployment | Depending on the requirements, the deployment phase can be as simple as generating a report or as complex as implementing a repeatable data mining process. In many cases it will be the customer, not the data analyst, who will carry out the deployment steps. However, even if the data analyst will not carry out the deployment, it is important for the customer to understand up front the actions which will need to be carried out in order to actually make use of the created models. |

TABLE II.        OBJECTIVES & BENEFITS OF CRISP-DM

| Sl. No | Objective |
|---|---|
| 1 | To ensure high quality of knowledge discovery from project results |
| 2 | To reduce skills required for knowledge discovery |
| 3 | To Reduce Costs and Time |
| 4 | Stable Model development across varying applications & generic in purpose |
| 5 | Should be robust and withstand changes in Environment |
| 6 | DM Modeling need to be Tool & Technique Independent |
| 7 | DM Methodology should be Tool Supportable |
| 8 | DM Methodology need to support documentation of projects |
| 9 | CRISP is designed to capture experience for reuse |
| 10 | CRSIP methodology supports knowledge transfer and training |

We have used tools available in SAP Business Information Warehouse (BIW) to handle outliers, missing, inconsistent and duplicate values in the source data [8].  The commonly sought supplementary value from this DM Model includes preventing fraud, giving marketing advice, seeking profitable customers, predicting sales and inventory and correcting data during bulk loading of the database, also known as the Extract-Transform-Load operation (ETL) [9]. Motivation for using DM comes from the value it gives over the competitors and it almost every time reduces costs, i.e. saves money, if the process is successful (Lukawiecki, 2008) [10].

Two high-level DM goals are prediction and description. The first one tries to find patterns to predict the value or behavior of some entity in the future and the second one tries to find patterns describing the values or behavior in a form understandable to humans. These high-level goals are pursued through several DM methods, for example classification, regression, clustering, summarization, dependency modeling and change, as well as deviation detection [11]. Each method has many algorithms that can be used to reach the goal, but some algorithms suit some problem areas better than others. (Fayyad et al., 1996c). The ETL map developed in SAP BIW workbench for InfoCube Updation is displayed below. This InfoCube is the data source for DM Model, to be developed later in APD (Analysis Process Designer) workbench of SAP.

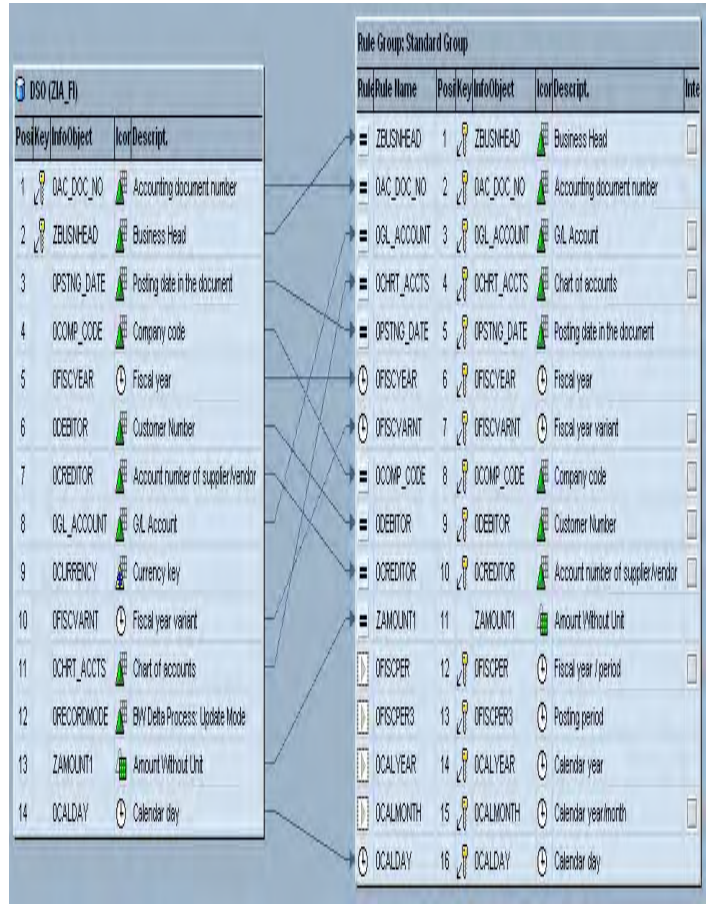

Figure 4.    ETL Map for InfoCube Loading in SAP BIW

APD is a workbench that is used to visualize, transform and deploy data from Business warehouse [12] [13]. APD tool supports KDD process where we can merge and manipulate data sources for complex data mining requirements

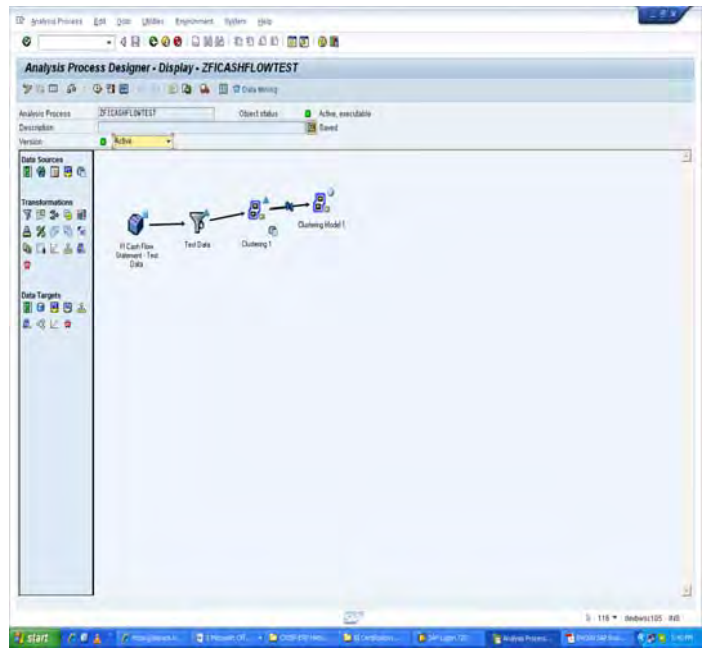

Figure 5.    SAP APD Workbench for DM Modeling & Visualization

## V. DATA MINING MODELING & IMPLEMENTATION

The specific steps with respect to CRISP-DM are outlined below.

### A. Business Understanding

After discussions with Finance function module leads , it is decided to mine data pertaining to vendors, Receipt & Payment functions, G/L Accounts for each posting date and corresponding amounts posted thereof. It is also decided to develop a Mining model to predict likely G/L accounts for certain vendors which exhibit similar attributes. From this information we can estimate amounts likely to be debited or credited on each posting date against each vendor. This information can be further utilized in Operations & Investment sub functions.

### B. Data Understanding

Transactional data generated in SAP ECC data sources is pulled into SAP Business warehouse by means of custom defined data extractors [13]. The extractor contains Meta information about source data and is capable of supporting delta mechanism, to update BIW with changed attribute values. Data quality issues are handled in BIW workbench by assigning relevant values for attributes and the considered attributes on DM Model are shown in table III below

### C. Data Preparation

Extracted source data enters PSA (Persistent Staging Area) layer of SAP BIW. Here we can edit any record and fix errors, if any. The granularity level at PSA layer is same as what we extracted from source systems. After cleansing data we loaded it into a DSO (Data Store Object) [13] [16]. The data in DSO is stored in a flat file which can be further used to load SAP InfoCube or InfoProvider. The Dimensions, Characteristics and Key Figure associated with our target InfoCube are shown in Table IV below

### D. Modeling

Choosing the right Data Mining Model is a complex task and should take into account the precise objectives for designing the analysis process. With SAP APD tool we can develop Mining Models using DM Algorithms like – Clustering, Classification, Association Analysis and Regression Analysis. A brief gist on these DM algorithms [16] and reasons for choosing a particular algorithm are given below.

1) *Association Rule Mining (ARM) or Association Analysis* is a Data Mining model designed to determine associations between different events.  The purpose of association analysis is to find patterns, in particular in business processes, and to formulate suitable rules, such as "If a customer buys product A, that customer also buys products B and C". Association Analysis Models are used to identify cross-selling opportunities for different products.  Most of the research focuses on the frequent itemset mining subproblem, i.e., finding all frequent itemsets each occurring at more than a minimum frequency (min. support) among all transactions [6]. Well-known sequential algorithms include Apriori [7], Eclat [8], FP-growth [9], and D-CLUB [10] [16]. For Cash Flow data analysis, ARM model is not found suitable and hence the analysis results are not discussed.

2) *Classification Algorithms:* These methods try to categorize items to predefined target classes with the help of some algorithm. The building of classification model includes training data set with known, discrete target classes, which means that the classification results are always discrete [14]. Classification targets vary from binary to multiclass attributes and the models try to predict which target class is correct with the help of descriptive relationships from the input attributes. Data classification is a two phase process in which first step is the training phase where the classifier algorithm builds classifier with the training set of tuples and the second phase is classification phase where the model is used for classification and its performance is analyzed with the testing set of tuples [15]. Decision Tree is a Classification scheme which generates a tree and a set of rules, representing the model of different classes, from a given data set [16]. The set of records available for developing classification methods is generally divided into two disjoint subsets – Training set & Test set. The former is used for deriving the classifier, while the later is used to measure the accuracy of classifier. Also, the accuracy of the classifier is determined by the percentage of test examples that are correctly classified. Decision Tree Results from our study are presented in next section.

3) *Regression Algorithms:* A Regression model predicts the value of a numerical data field, this is the target field, in a given data record from the known values of other data fields of the same record. The known values of other data fields are called input data fields or explanatory data fields. They can be numerical or categorical. The predicted value might not be identical to any value contained in the data used to build the model [17]. A regression model is created and trained based on known data sets of data records whose target field values are known. You can apply the trained model to known or to unknown data. In unknown data, the values of the input fields are known; however, the value of the target field is not known [11]. A simple case of linear regression, where the sum of squared errors is minimized when –

$$w = \sum \frac{x_i y_i}{\sum x_i^2}$$

The maximum likelihood model is $out(x) = wx$ , which is used for prediction

4) *Clustering Algorithms:* Clustering in data mining is a discovery process that groups a set of data such that the intracluster similarity is maximized and the intercluster similarity is minimized. When there is no specified class, clustering is used to group items that seem to fall naturally together [16]. For our study clustering algorithms are found most suitable to identify valuable clusters for analysis. Clustering algorithms are mainly of 3 types namely – Hierarchical, Partition based and Density based Methods [18]. The important algorithms under each of these methods are discussed here.

a) *Hierarchical Methods*: This method groups data objects into a tree of clusters. This method can be further classified as either *agglomerative* or *divisive*, depending on whether the hierarchical decomposition is formed in a bottom-up (merging) or top-down (splitting) fashion [19]. This method examines all the clusters present so far at each stage of merging; the clustering methods we examined work incrementally, instance by instance [20]. At any stage the clustering forms a tree with instances at the leaves and a root node that represents the entire dataset. In the beginning the tree consists of the root alone. Instances are added one by one, and the tree is updated appropriately at each stage. The important algorithms in hierarchical clustering are -

    i.     *Agglomerative Clustering* (*Bottom-up*): It measures distance between any 2 clusters and treats any instance as a cluster in its own right, then finds the two closest clusters, merges them, and keep on doing this until only one cluster is left. The record of merging forms a hierarchical clustering structure—a binary dendrogram [21]

    ii.    *Cobweb Algorithm*: It always compares the best host, adding a new leaf, merging the two best hosts, and splitting the best host when considering where to place a new instance

b) *Partition based Method:* The clusters are formed to optimize an objective partitioning criterion, such as a dissimilarity function based on distance, so that the objects within a cluster are "similar," whereas the objects of different clusters are "dissimilar" in terms of the data set attributes. Given D, a data set of n objects, and k, the number of clusters to form, a partitioning algorithm organizes the objects into k partitions (k ≤ n), where each partition represents a cluster [22]. Important partition clustering algorithms are K-Means & EM (Expectation-Maximization)

c) *Density based Methods*: This discovers clusters with arbitrary shapes & typically regard clusters as dense regions of objects in the data space that are separated by regions of low density (representing noise) [23] [24]. DBSCAN grows clusters according to a density-based connectivity analysis. OPTICS extends DBSCAN to produce a *cluster ordering* obtained from a wide range of parameter settings. DENCLUE clusters objects based on a set of density distribution functions [7] [8].

*E.   Evaluation*

For evaluating Cluster, Classification and Regression models we have split input data for training and testing purposes. The splitting criterion is, 66% of data is used for training and the rest for testing the model results. Cluster model has formed 10clusters and used binning intervals to group values together on output visualization [21]. Many *What-If* scenarios were run on these models and accuracy is found to be within desired range. The predicted attribute values are validated against known results and accordingly decision model input parameters like number of leaf nodes, stooping criteria are adjusted, to achieve desired model accuracy [24]. Care is taken to avoid overfitting of data so that the model can accommodate variations in input data [25].

*F.   Deployment*

The results of SAP Data Mining models can be accessed by all the concerned in following ways [13] [16]–

    i.     Model results can be written on to a flat file and published on enterprise portals of SAP i.e. at people integration layer

    ii.    Model results are graphically visualized and desired result charts are broadcasted to all stakeholders

    iii.   Model results in text form are fed back into Enterprise Reports for decision making at all levels

## VI.   RESULTS & DISCUSSION

*Decision Tree* is a classification approach that uses the value of input variables to predict the value of a categorical variable. In SAP Decision Tree Model we have defined G/L Account as predicted variable [20]. Field DT_Pred_Val002 in table V represents value predicted by DM Model for G/L Account. Field DT_Pred_Node002 denotes node value in Decision tree & Field DT_Pred_Prob002 represents probability of prediction accuracy. Obviously probability value of 1.0 gives absolute accuracy of predicted result but results are also fairly accurate with probability values like 0.2, 0.48,0.78 etc as shown table V below.

*Parameters used for Cluster model* are – Binning Interval – 10; Default weight on variable – 1.0. In Table VI below, Field CL_Pred_Cluster002 represents predicted cluster for input data attributes. From Overall Influence Chart in Figure 7 below, Cluster No 6 has most number of records and could be used for prediction of unknown data values [16].

*For Regression Model*, field SC_Score002 represents predicted score for attribute *amount*

TABLE III.        SOURCE FIELD ATTRIBUTES IN SAP SYSTEM FOR CASH FLOW STATEMENT

| Field | Description | SAP Data Type | Length | Default Value |
|---|---|---|---|---|
| BUKRS | Company Code | Char | 4 | 1000 |
| SAKNR | G/L Account | Char | 10 | No Default Value |
| BUDAT | Posting Date | Dats | 8 | No Default Value |
| BIZ_HEAD | Business Head | Char | 10 | OPER |
| WRBTR | Amount in LC | Curr | 13 | No Default Value |
| BELNR | Document Number | Char | 10 | No Default Value |
| GJAHR | Fiscal Year | Numc | 4 | 2012 |
| LIFNR | Vendor | Char | 10 | No Default Value |
| KUNNR | Customer | Char | 10 | No Default Value |
| WAERS | Currency | Cuky | 5 | INR |

TABLE IV.        DIMENSIONS, CHARACTERISTICS AND KEY FIGURES OF INFOCUBE

| Dimension | Characteristics | Technical Name | Data Type | Length |
|---|---|---|---|---|
| Business Head (ZFI_CASH1) | Business Head | ZBUSNHEAD | Char | 10 |
| | Accounting Document No | 0AC_DOC_NO | Char | 10 |
| | Company Code | 0COMP_CODE | Char | 04 |
| GL Account (ZFI_CASH2) | G/L Account | 0GL_ACCOUNT | Char | 10 |
| | Chart of Accounts | 0CHRT_ACCTS | Char | 04 |
| | Posting date in the Document | 0PSTNG_DATE | Dats | 08 |
| Customer (ZFI_CASH3) | Customer Number | 0DEBITOR | Char | 10 |
| | Account Number of Vendor | 0CREDITOR | Char | 10 |
| Key Figure | Amount Without Unit | ZAMOUNT1 | Curr | 09 |

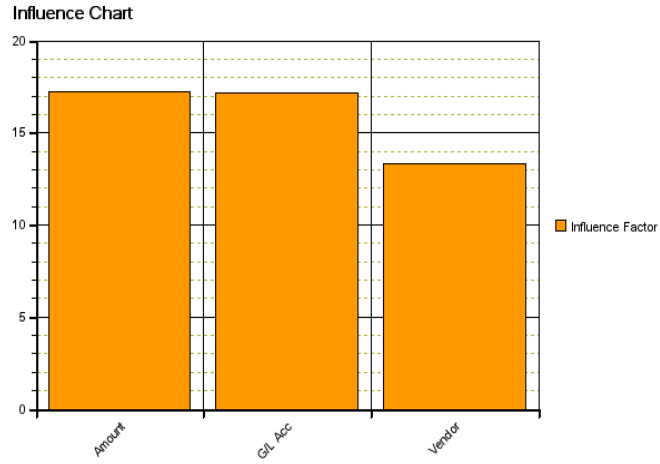

Figure 6.   Clustering Model: Relative Dominance of each attribute in Overall Influence Chart

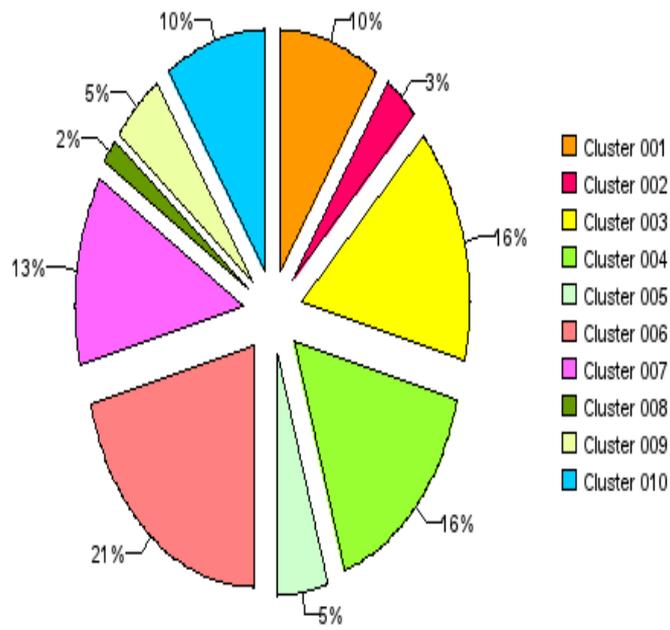

Figure 7.   Clustering Model Overall Influence Chart

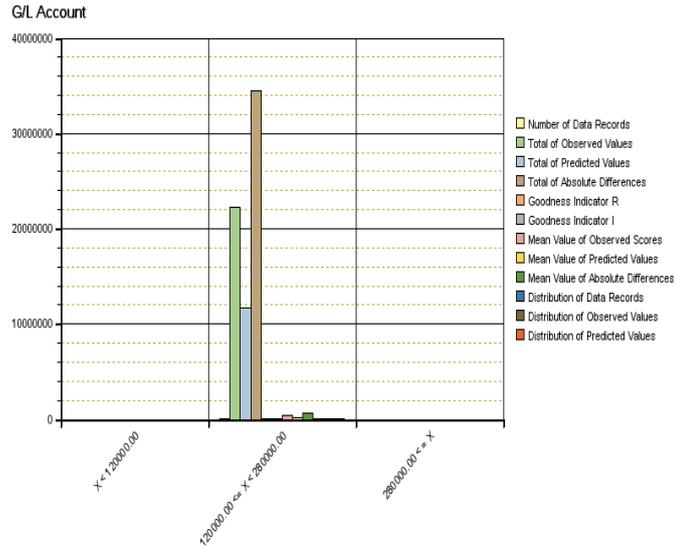

Figure 8.   Regression Model: GL Account Scoring Output

TABLE V.        DECISION TREE DM MODEL SAMPLE RESULTS FOR PREDICTED  VARIABLE

| 0CREDITOR | 0GL_ACCOUNT | 0PSTNG_DATE | ZAMOUNT1 | DT_PRED_NODE002 | DT_PRED_PROB002 | DT_PRED_VAL002 |
|---|---|---|---|---|---|---|
| 100888 | 272610 | 20120810 | 1000 | 0 | 0.2 | 181030 |
| 114033 | 259010 | 20120822 | 100000 | 11 | 1 | 259010 |
| 118203 | 259010 | 20120831 | 4000 | 11 | 0.2 | 181030 |
| 118203 | 259010 | 20120906 | 3000 | 11 | 0.2 | 181030 |
| 200710 | 181030 | 20120811 | 1853 | 4 | 0.78571 | 181030 |
| 201297 | 120012 | 20120817 | 10530 | 19 | 0.47826 | 120012 |
| 600010 | 181030 | 20120810 | 200 | 24 | 0.75 | 161011 |

TABLE VI.        CLUSTERING DATA MINING MODEL SAMPLE RESULTS

| 0AC_DOC_NO | 0CREDITOR | 0GL_ACCOUNT | 0PSTNG_DATE | ZAMOUNT1 | CL_PRED_CLUSTER002 |
|---|---|---|---|---|---|
| 1226000224 | 200031 | 250602 | 20120804 | 121212 | 2 |
| 1226000227 | 200031 | 250602 | 20120804 | 3000000 | 2 |
| 1226000228 | 200031 | 250602 | 20120804 | 3000000 | 2 |
| 1226000229 | 200031 | 250602 | 20120804 | 300000 | 2 |
| 1226000237 | 201402 | 181030 | 20120806 | 2000 | 5 |
| 1226000239 | 201196 | 250022 | 20120807 | 3191271 | 6 |

TABLE VII.        REGRESSION DATA MIMING MODEL SAMPLE RESULTS

| 0AC_DOC_NO | 0CREDITOR | 0GL_ACCOUNT | 0PSTNG_DATE | ZAMOUNT1 | SC_SCORE002 |
|---|---|---|---|---|---|
| 1226000224 | 200031 | 250602 | 20120804 | 121212 | 5.31E+04 |
| 1226000227 | 200031 | 250602 | 20120804 | 3000000 | 5.31E+04 |
| 1226000228 | 200031 | 250602 | 20120804 | 3000000 | 5.31E+04 |
| 1226000229 | 200031 | 250602 | 20120804 | 300000 | 5.31E+04 |
| 1226000230 | 200031 | 250602 | 20120804 | 3000000 | 5.31E+04 |

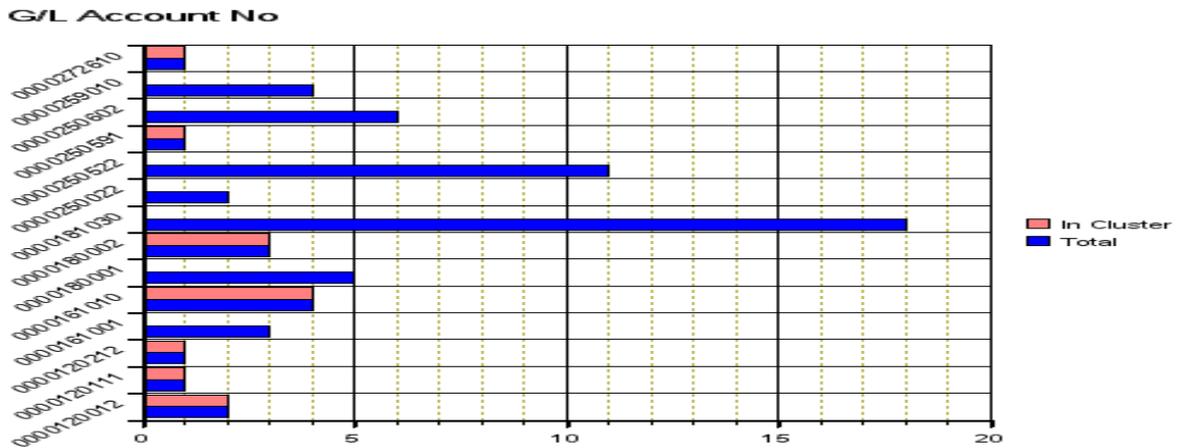

Figure 9.   Clustering Model for GL Account No Attribute Value Distribution Chart

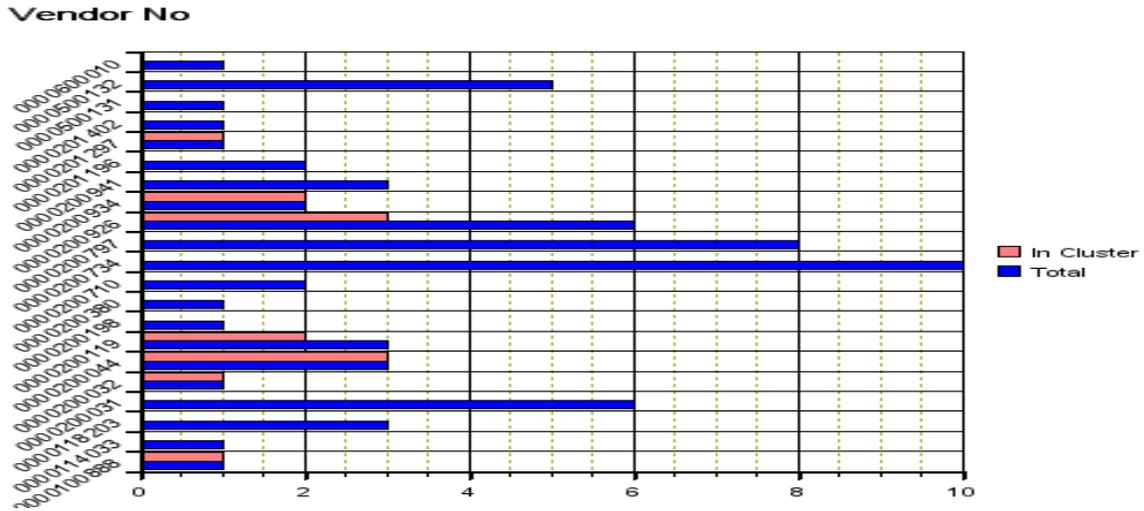

Figure 10.  Clustering Model for Venodr No Attribute Value Distribution Chart

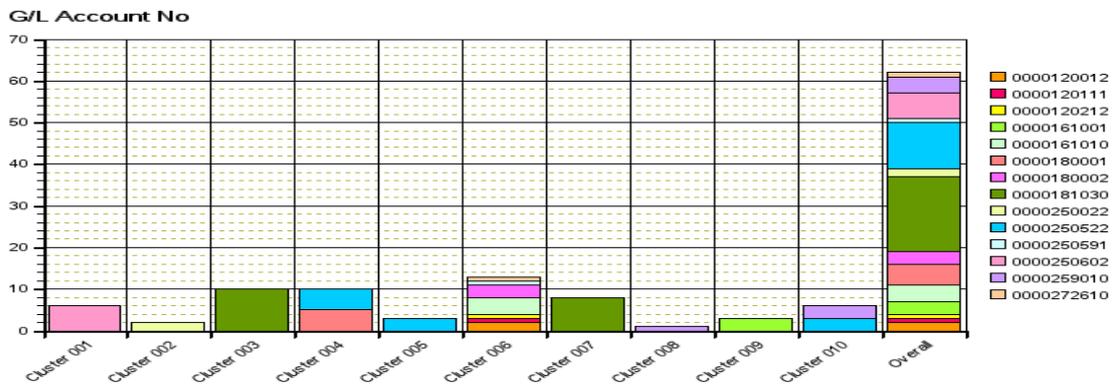

Figure 11.  Attribute Value Chart for G/L Account No in each Cluster

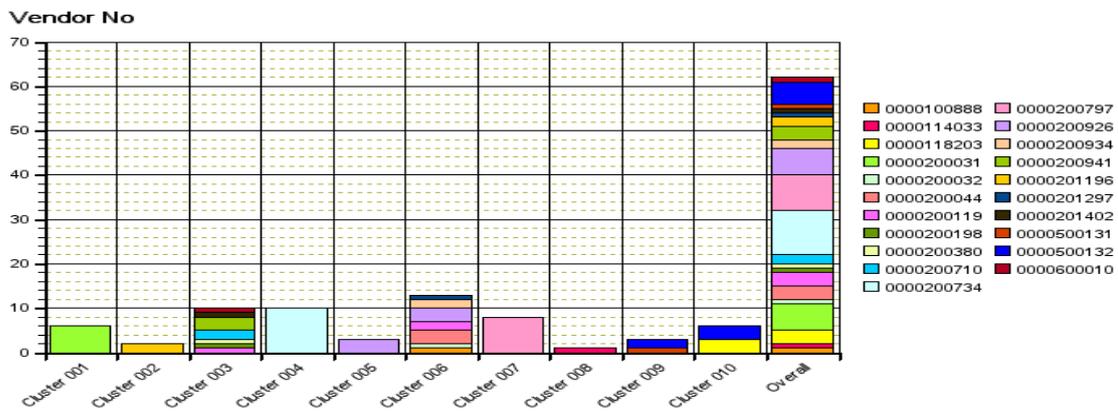

Figure 12.  Attribute Value Chart for Vendor No in each Cluster

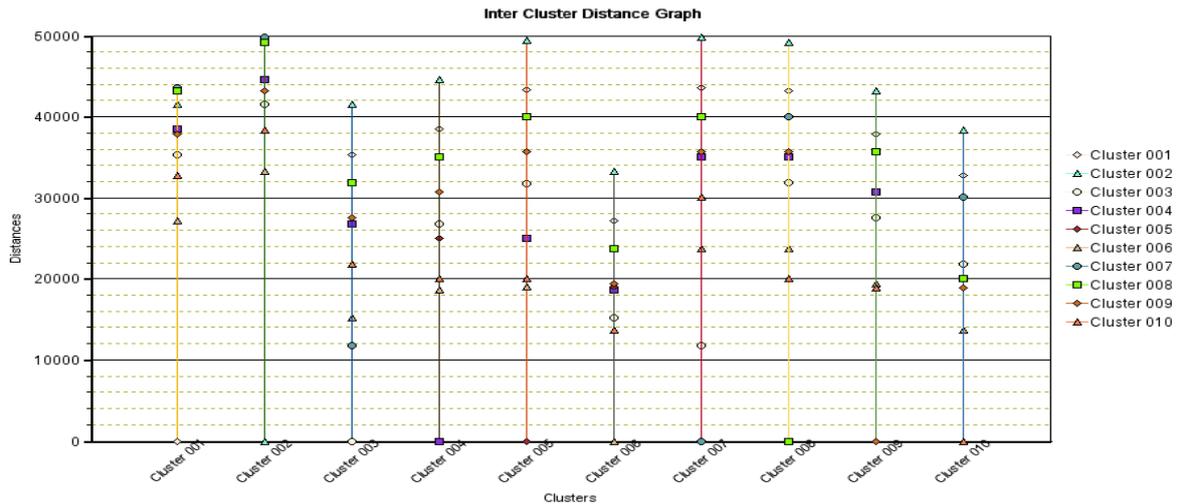

Figure 13. Inter Cluster Distance Graph

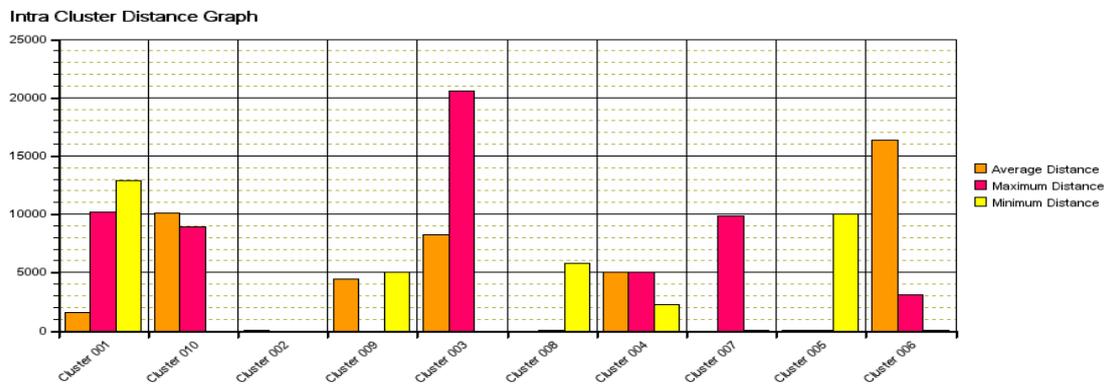

Figure 14. Intra Cluster Distance Graph

## VII. CONCLUSIONS

ERP systems has limited capabilities with regard to, analytics that can answer key business questions, Collective Intelligence and Actionable insight to quickly respond to market demands with appropriate decisions at each level. Business analysts rely heavily on query and reporting to provide them with the information they need to connect the dots between revenues and losses, products and profit ability, financial performance and market trends and so on. They need a comprehensive query and reporting capability that can tap knowledge from huge volumes of ERP systems and here CRISP-DM provides a methodology which can be adopted to support these requirements. With Data Mining Analytics, ERP Systems can facilitate exploration of all types of information from all angles to assess the current business situation, analyze facts and anticipate tactical and strategic implications with more advanced, predictive or what-if analysis. The result of a Data Mining project under CRSIP-DM pedagogy is not just Models but also findings that are important in meeting the objectives of the business or important in leading to new questions, lines of approach, or side effects. There are diverse subjects for future work and research, like mapping out more problem regions in ERP systems and descriptive attributes with CRSIP-DM by exploring more data sets.

### REFERENCES

[1] Óscar Marbán, Gonzalo Mariscal and Javier Segovia, "A Data Mining & Knowledge Discovery Process Model". In Data Mining and Knowledge Discovery in Real Life Applications, Book edited by: Julio Ponce and Adem Karahoca, , pp. 438-453, February 2009

[2] Lukasz Kurgan and Petr Musilek (2006), "A survey of Knowledge Discovery and Data Mining process models". The Knowledge Engineering Review. Volume 21 Issue 1, March 2006, pp 1 - 24, Cambridge University Press, New York, NY, USA

[3] Azevedo, A. and Santos, M. F. **(2008),**"KDD, SEMMA and CRISP-DM: a parallel overview". In Proceedings of the IADIS European Conference on Data Mining 2008, pp 182-185

[4] Sistla Hanumanth Sastry, Prof. M. S. Prasada Babu. " ERP implementation for Manufacturing Enterprises." International Journal of Advanced Research in Computer Science and Software Engineering (IJARCSSE), Vol. 3, Issue 4, pp 18-24, April 2013.

[5] Pete Chapman, Julian Clinton , Randy Kerber et al, "The CRISP-DM User Guide", 1999

[6] Pete Chapman, Julian "Glintin" Clinton, Randy Kerber, Thomas Khabaza, Thomas Reinartz, Colin Shearer, and Rüdiger Wirth , "CRISP-DM 1.0 Step-by-step data mining guide", 2000

[7] Oscar Marban, Gonzlo Mariscal et al;" A Data Mining & Knowledge Discovery Model", I-Tech, Vienna, Austria,pp 438, February 2009

[8]  Sistla Hanumanth Sastry,  Prof. M. S. Prasada Babu.  "Implementing a successful Business Intelligence framework for Enterprises." Journal of Global Research in Computer Science (JGRCS), Vol. 4, No. 3 (2013): pp. 55-59. April 2013.

[9]  Han., M.Kamber , "Introduction to Data Mining" Morgan Kaufman Publishers, 2006 ;pp: 429-462

[10]  Lior Rokach, Oded Maimon, "DATA MINING AND KNOWLEDGE DISCOVERY HANDBOOK" 2010, Springer USA; pp 322-350

[11]  Margaret H. Dunham, "Data Mining: Introductory and Advanced Topics", 2003,Prentice Hall of India; pp: 135-162

[12]  Johannes Grabmeier, Andreas Rudolph Data Mining and Knowledge Discovery," Techniques of Clustering Algorithms in Data Mining";Volume: 6, Issue: 1996, Publisher: Springer, Pages: 303-360

[13]  Sistla Hanumanth Sastry,  Prof.  M. S.Prasada Babu,  "Cluster Analysis of Material Stock Data of Enterprises",  International Journal of Computer Information Systems (IJCIS),  Vol. 6,  Issue 6,  pp. 8-19,  June 2013.

[14]  Pang-Ning Tan, Michael Steinbach, Vipin Kumar, "Introduction to Data Mining", March 2006.,published by Addision-Wesley , PP 330-340

[15]  Hand D., Mannila, H. and Smyth, P.,  "Principles of Data Mining", 2001,Prentice Hall of India; pp: 292-305

[16]  Sistla  Hanumanth Sastry,  Prof. M. S. Prasada Babu, " Analysis of Enterprise Material Procurement Leadtime using Techniques of Data Mining.",  International Journal of Advanced Research in Computer Science (IJARCS), Vol. 4,  Issue 4,  pp. 288-301,  April 2013.

[17]  Mark Hall ,Ian Witten , Eibe Frank, "Data Mining Practical Machine Learning Tools & Techniques", January 2011 ; Morgan Kaufmann Publishers; pp: 278-315

[18]  Galit Shmueli, Nitin R. Patel ,Peter C. Bruce; "Data Mining for BusinessIntelligence", 2007, John Wiley & Sons; pp: 220-237

[19]  Dorian Pyle, "Data Preparation for Data Mining", Morgan Kaufmann Publishers,1999, San Francisco, USA; pp: 100-132

[20]  Boudaillier.E et al;"Interactive Interpretation of Hierarchical Clustering"; Principles of Data Mining and Knowledge Discovery: Proceedings of First European Symposium, PKDD'97, Trondheim, Norway June 24-27,1997; pp: 280-288

[21]  Ester, M., Kriegel, H.-P., Sander, J. & Xiaowei, X., 1996. "A Density-Based Algorithm for Discovering Clusters in Large Spatial Databases with Noise". In *Proc. of 2nd International Conference on Knowledge Discovery and Data Mining (KDD '96).*, 1996. AAAI Press.

[22]  Fayyad, U., Piatetsky-Shapiro, G. & Smyth, P., 1996b. "From Data Mining to Knowledge Discovery in Databases". *AI Magazine*, 17, pp.37-54.

[23]  Kurgan, L.A. & Musilek, P., 2006. A survey of Knowledge Discovery and Data Mining process models. The Knowledge Engineering Review, 21(1), pp.1-24.

[24]  Sistla Hanumanth Sastry,  Prof.  M. S.Prasada Babu,  "Performance evaluation of clustering Algorithms", International Journal of Computational Science and Information Technology,  Unpublished.

[25]  Shtub, A., 2002. "*Enterprise Resource Planning (ERP) : the dynamics of operations management*". Boston: Kluwer Academic

**AUTHORS PROFILE**

S.Hanumanth Sastry (Corresponding Author) Senior Manager (ERP) has implemented SAP-BI Solutions for Steel Industry. He holds M.Tech (Computer Science) from NIELIT, New Delhi and MBA (Operations Management) from IGNOU, New Delhi. His research interests include ERP systems, Data Mining, Business Intelligence and Corporate Performance Management. He is pursuing PhD (Computer Science) from Andhra University, Visakhapatnam (INDIA).

Prof. M.S. Prasad Babu obtained his Ph.D. degree from Andhra University in 1986. He was the Head of the Department of the Department of Computer Science & Systems Engineering, Andhra University from 2006-09. Presently he is the Chairman, Board of Studies of Computer Science & Systems Engineering. He received the ISCA Young Scientist Award at the73rd Indian Science Congress in 1986.